**Biomagnetic signals recorded during transcranial magnetic stimulation (TMS)-evoked peripheral muscular activity**

Geoffrey Z. Iwata*[1], Yinan Hu*[1,a)], Tilmann Sander[2], Muthuraman Muthuraman[3,b)], Venkata Chaitanya Chirumamilla[3], Sergiu Groppa[3], Dmitry Budker[1,4,5] and Arne Wickenbrock[1,4]

1) Institut für Physik, Johannes Gutenberg-Universität Mainz, 55128 Mainz, Germany.
2) PhysikalischTechnische Bundesanstalt, Berlin, Germany.
3) Movement Disorders and Neurostimulation, Biomedical Statistics and Multimodal Signal Processing Unit, Department of Neurology, University Medical Center of the Johannes Gutenberg-University Mainz, Mainz-55131, Germany.
4) Helmholtz Institute Mainz, GSI Helmholtzzentrum für Schwerionenforschung, 55099 Mainz, Germany.
5) Department of Physics, University of California, Berkeley, CA 94720-7300, USA
*These authors contributed equally to this work
a) Corresponding author; Electronic mail: yinanhu1@uni-mainz.de
b) Corresponding author; Electronic mail: mmuthura@uni-mainz.de

*Abstract— Objective:* We present magnetomyograms (MMG) of TMS-evoked movement in a human hand, together with a simultaneous surface electromyograph (EMG) and electroencephalograph (EEG) data. *Approach:* We combined TMS with non-contact magnetic detection of TMS-evoked muscle activity in peripheral limbs to explore a new diagnostic modality that enhances the utility of TMS as a clinical tool by leveraging technological advances in magnetometry. We recorded measurements in a regular hospital room using an array of optically pumped magnetometers (OPM) inside a portable shield that encompasses only the forearm and hand of the subject. *Main Results:* The biomagnetic signals recorded in the MMG provide detailed spatial and temporal information that is complementary to that of the electric signal channels. Moreover, we identify features in the magnetic recording beyond those of the EMG. *Significance:* These results validate the viability of MMG recording with a compact OPM based setup in small-sized magnetic shielding, and provide proof-of-principle for a non-contact data channel for detection and analysis of TMS-evoked muscle activity from peripheral limbs.

## I. INTRODUCTION

The central nervous system of the human body forms a critical signaling network that controls over 200 muscles[1]. Developing new technologies that aid in understanding and measuring the innervation patterns and muscle activity controlled by this network is crucial for the advancement of research, diagnosis and treatment of motor-system diseases like Parkinson's disease and amyotrophic lateral sclerosis (ALS)[2,3]. One such technology is transcranial magnetic stimulation (TMS), which has recently gained widespread use for research and diagnosis of various neuropsychiatric disorders[4–7]. In TMS, a strong magnetic field pulse is applied to a specific cortical area. When applied to the motor cortex, TMS results in an evoked muscle response in the form of a 'twitch'[8–10]. TMS offers a safe, controlled, and non-invasive method to investigate the entire motor pathway from motor cortex to muscle, making it an ideal platform for studying central motor conduction activity[11-13]. The TMS-evoked muscle and nerve responses in descending motor waves can be recorded with electrophysiological measurement techniques such as electromyography[14] (EMG), which is considered a gold-standard tool. The electromyography signal from a muscle stimulated via TMS is known as a motor evoked potential (MEP)[12]. Magnetic signals accompany electrophysiological signals and can provide clinically relevant information about innervation and muscle activity that is spatially and temporally well resolved[15,16]. A major challenge is to combine the utility of TMS with the stringent operational requirements of state-of-the-art magnetometers. Here, we overcome those challenges and record magnetic fields from TMS-evoked muscle activity acquired in a regular hospital room using optically pumped magnetometers and small-sized magnetic shielding.

Biomagnetic measurements can offer complementary data in TMS-EMG experiments aiming to measure evoked muscle activity, since a magnetic response will accompany the changing electric field. Surface EMG records electrical potential differences that arise due to electromagnetic activity associated with so called motor-unit action potentials (MUAPs), which are a summation of individual muscle action potentials that propagate along a single contracted muscle fiber[14]. The surface EMG does not measure the direct action-potential in the muscle, but rather the associated ensemble electromagnetic activity that reaches the skin at a specific moment in time. Since electric fields in the body are affected by the conductivity of different tissues and specific skin conditions, it can be challenging to recover the exact origin of the EMG signal without complex and careful electrode placing and analysis of the specific physiological conditions[17].

Meanwhile, magnetic fields arise from the composite electrical activity within the body and thus also require detailed analysis to recover source information. These fields convey information from both the primary MUAPs, as well as the secondary propagation of electrical activity through the surrounding biomass. Despite this complication, detailed array measurements of the field can also be used to locate the primary sources with excellent agreement with established EMG localization techniques[18]. In this manuscript, we refer to the recorded magnetic field signal that arises from a MUAP as a motor evoked field (MEF). Importantly, since the relative magnetic permeability of human tissue is close to unity, the magnetic fields from MUAPs are directly related to the electro-chemical activity within muscles, unaffected by specific conditions of the surrounding tissue[19]. Crucially, they do not rely on a sensor-skin connection. Therefore, while electrical and magnetic signals originate from a single event, the difference in how they are communicated to a sensor means that they can validate each other and provide complementary information about the system under study. In combination with TMS, these magnetic signals can elucidate the proper functioning and response of the muscular and central nervous systems[20].

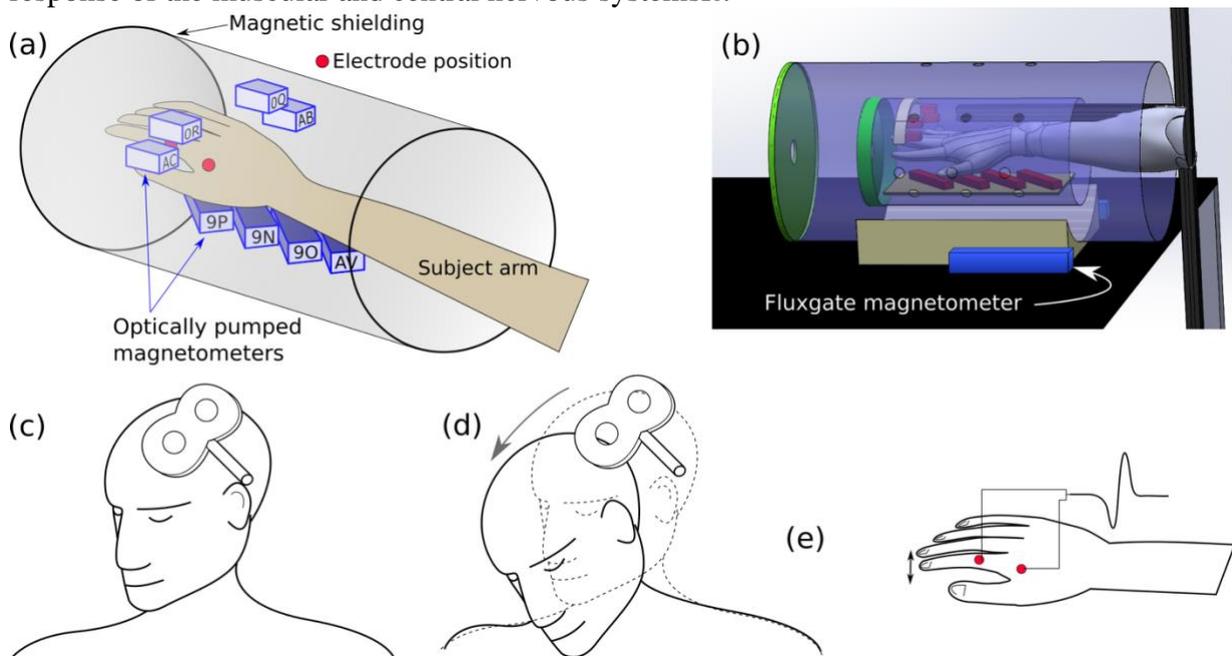

**FIG. 1:** Experimental setup. (a) Schematic of a subject's hand within the innermost magnetic shield layer. Sensor positions of the magnetometers and electrodes are indicated. Not indicated

are mounting/supporting elements or wires. (b) Rendering of experimental setup showing subject's hand within the innermost magnetic shield layer and outermost shielding layer. Optically pumped magnetometer (OPM) positions are shown in red. (c) Schematic of a participant's head with a TMS coil positioned above the motor cortex. (d) Control measurement to identify magnetic artifacts in sensor output arising from the TMS pulse. When the participant moves their head down, the TMS coil does not stimulate the motor cortex, but the magnetic artifact at the sensors is the same as in (c). (e) Motor evoked potentials (MEPs) and motor evoked fields (MEFs) were recorded from the right first dorsal interosseus (FDI) muscle during TMS (ground electrode is on the index finger). The stimulus results in a lateral `twitch' of the right index finger.

Since the detection of magnetic fields does not require physical contact, a magnetic measurement of muscle activity, or magnetomyography (MMG)[21], is a correspondingly non-contact technique. These aspects make MMG an attractive tool for complementing EMG, since magnetic signals can cross-validate electrophysiological measurements by decoupling signal strength from changes in systematic experimental conditions, such as electrode-skin contact for electrodes and background magnetic field changes in magnetometers. TMS provides ideal conditions for collecting data from these different sensing modalities. Because TMS is a repeatable and controlled stimulation, measurements can be triggered and averaged with high accuracy in timing, improving the signal-to-noise ratio and repeatability of biomagnetic signals. As a result, while information about individual trials is diminished, persistent features across repetitive stimulations can be analyzed in detail.

Despite the apparent motivations for magneto-physiological measurements, the very small signal size (<10 pT) has limited widespread adoption of biomagnetic measurements as a routine clinical measurement, since this regime of sensitivity has been limited to SQUIDs[22] (superconducting quantum interference devices), which require cryogenic cooling. As a result, while SQUIDs have been for decades used to detect biomagnetic signals[15,23–27], including those arising from periphery limbs, the associated measurement systems are bulky, expensive, and ill-suited to the different geometries of various body parts, limiting these systems' practical utility. Recent developments in atomic magnetometry have led to high-sensitivity devices known as optically pumped magnetometers (OPMs)[28] that are uncooled, centimeter-scale, and relatively low cost – characteristics necessary to make magneto-physiological measurements an accessible diagnostic tool. Additionally, OPMs have opportunities and applications in wearable compact devices with wide potential outside of clinical use[29]. For these reasons, OPMs have recently generated broad research interest as a viable alternative to SQUIDs in measuring weak biomagnetic signals.

Current OPM technology mandates heating the sensor, resulting in surface temperatures of around 40$_o$C, and requires the background magnetic field to be below ~50 nT – well below the Earth's magnetic field and typical noise sources (line noise, equipment noise, elevators, cars, etc.). To achieve this precondition, previous studies with OPMs have typically employed magnetically shielded rooms, which are incompatible with the intense fields produced by TMS.

In this work, we leverage the advances in OPM technology to enhance the diagnostic utility of TMS. Several recent studies have shown that OPMs can detect electrically stimulated muscle activity in the hand while in a magnetically shielded environment[30,31]. Our study represents several key advances. First, in using TMS, the evoked muscle activity in this study arises from signals that originate with magnetic stimulation at the motor cortex, and therefore involves the entire motor pathway, in contrast to electrical stimulation of proximal nerves[31]. Second, we incorporate simultaneous electroencephalography (EEG) during measurements, in addition to

EMG. Thus, we extend magnetic measurements to an established routine to study repetitive TMS evoked activity recorded with EEG and EMG[32-34]. Finally, we achieve the above stated goals in a regular hospital examination room, by using a portable magnetic shield that only encompasses the arm of the subject.

With this unique setup, we recorded biomagnetic signals with features that complement and are validated by EMG measurements, and furthermore, the magnetometers detect signals from parts of the hand that were not covered by the EMG electrodes.

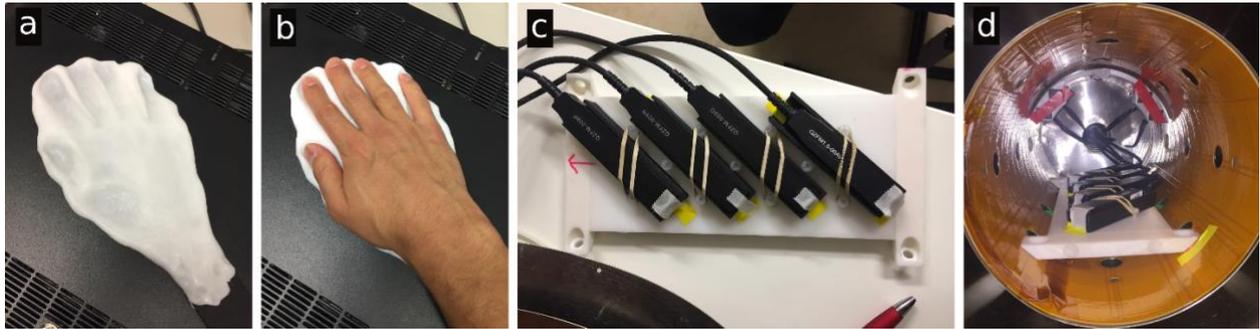

**FIG. 2:** (a)Thermo-plastic hand molds made for each subject to rest their forearm and hand on during the measurement. The plastic is molded around an aluminum support for the forearm. The mold section for the index finger is widened to allow for evoked motion due to the TMS. (b) Hand mold with a hand. Using flexible VELCRO® strips, the whole mold is suspended from an aluminum strut that extends into the magnetic shield. (c) Four commercial OPM sensors (QuSpin) arranged on a plastic board that is fit to the magnetic shield. (d) View inside the innermost shielding layer, with the eight OPMs seen (four more are behind the red tape at the top). Subjects place their forearm on the hand mount, which is then maneuvered into this magnetic shield. In contrast to a magnetically shielded room typically required for sensitive biomagnetic measurements, our setup avoids potentially claustrophobic conditions.

## II. METHODS
### A. Subjects
All measurements were repeated for four subjects in total, between ages 26 and 40, who volunteered for the study. All subjects are right-handed and have no somatic diseases or any mental or neurological diseases with confirmed diagnoses. Written informed consent in accordance with the Declaration of Helsinki was obtained from all subjects before participation in this study, which was approved by the Ethics Committee of the State Medical Association of Rhineland-Palatinate. Written informed consent was also obtained from all subjects to publish data/images relating to the experiment in an online open-access publication.

### B. Experimental procedure
The OPM and EMG electrode configurations within the shield are shown in Figure 1a-b. The measurement preparation time, including control measurements, takes less than 30 minutes. The biomagnetic signal is recorded with an array of four OPMs below the hand and an additional four above the hand, while the EMG and EEG are simultaneously recorded to correlate and provide reference for the signals.

For each participant, the three sensing systems (magnetomyography magnetometers, EMG surface electrodes, and EEG electrodes) were prepared and tested individually with the data-acquisition system. The EEG and EMG were recorded using a 256-channel EGI (Electrical

Geodesics, Inc.) EEG system and synchronized with OPM data using a trigger signal from the TMS pulse. The subject's hand was positioned inside the shield, and the TMS coil was positioned over the subject's left M1 region (Fig. 1c). The stimulation was applied at different frequencies 0.5 Hz, 3 Hz, and 9 Hz for a maximum of up to 3 mins.

The TMS pulse results in a ≈1.4 T field on the motor cortex of the participant, which is less than a meter away from the EMG and MMG sensor positions. The sensors record a magnetic artifact arising from the TMS pulse, which consists of a bi-phasic pulse lasting approximately 300 μs. To identify and isolate this artifact, a control measurement was performed in which each participant moved their head down (Fig. 1d) and data were taken for the same stimulation described above. Since the high-intensity and rapidly changing region of the magnetic field from the TMS is highly localized, the induced electrical field and resulting brain activity are also limited to a small volume. Therefore, the participants' change in head position results in the absence of a discernible evoked effect, but the magnetic artifact (from the TMS pulse) at the sensor is the same as in the experimental conditions. For this control measurement, no MEP is observed on the EMG, and a lack of finger 'twitch' was confirmed using a camera aimed through an access port of the shield. These measurements showed that the artifact lasts up to 15 ms on the averaged OPM signal – longer than the true pulse due to low-pass filtering in the sensor hardware. A small timing jitter in the system leads to artifact signal reduction in averaging. Since the jitter (<100 μs) is much smaller than the time-scale of signals of interest (>1 ms), this effect does not diminish signal amplitudes.

## C. Transcranial magnetic stimulation (TMS)

To administer TMS, a stimulation coil is placed over of the target area of a participant's scalp, and an electrical current running through the coil results in a region of intense magnetic field (up to 1.4 T) within the participant's brain. This pulsed magnetic field induces a secondary electrical current within cortical tissue, which, if within the motor cortex, may result in muscular activation[33].

The Magstim Super Rapid 2 stimulator (Magstim, UK) with a figure-of-eight coil and internal wing diameter of 70 mm was used. The TMS pulse had a bi-phasic waveform and was applied at the left primary motor cortex (M1) with an intensity of 110% of the subjects resting motor threshold (RMT) (Fig. 1c). The RMT was determined as the minimum stimulus intensity required to elicit motor evoked potentials of amplitude 50 μV in 5 out of 10 consecutive trials at rest in the contralateral first dorsal interosseous (FDI) muscle (Fig. 1e)[12].

## D. Optically pumped magnetometry in a portable shield

Detection of biomagnetic signals requires magnetic sensitivities better than 10 pT/√Hz. The commercially available OPMs (QuSpin) used in this work can achieve a noise floor of 15 fT/√Hz with a bandwidth between 1 - 100 Hz. These sensors operate by optically probing the zero-field resonance of spin-polarized rubidium atoms, which is highly sensitive to small magnetic fields[35]. Eight OPM sensors were used and each sensor has two magnetically sensitive axes, resulting in a total of 16 magnetic sensor channels.

The main drawback of this magnetometry approach is limited dynamic range, requiring a magnetically compensated or shielded background environment in order to reach the sensitivity limits, especially when considering a magnetically hostile hospital setting. Most previous human biomagnetic measurements using OPMs[29-31] were conducted in magnetically shielded rooms (MSRs) which typically have residual fields of <10 nT, magnetic gradients on the order of 1 nT/m[36], and enough space to comfortably accommodate a subject. These characteristics constitute an appropriate working environment for OPMs, allowing low-noise measurements and some freedom to move the sensors by 1-2 cm[37]. However, MSRs are expensive and not portable, which ultimately restricts the OPM technology to similar limitations as SQUID devices. Furthermore, the isolated MSR environment can be unsuitable for subjects to remain inside for long measurement times. Importantly, the large magnetic field generated by TMS could magnetize and negatively affect the shielding.

To circumvent these practical issues associated with MSR, we instead use a small-sized shield that encompasses only the body part relevant to the measurement. Since we are measuring nerve and muscle activity in the hand, the arm of the subject is placed inside a commercially available four-layer cylindrical shield (Twinleaf MS-2) with one set of end-caps removed. The missing end-cap compromises the DC shielding by about a factor of 10 within the sensor region, however, DC magnetic field offsets (<50 nT at sensor positions) arising in the shield can be compensated with the sensors' compensation coils. Nevertheless, the open-shield modification makes the low-field region susceptible to environmental magnetic noise, therefore the ability to average over multiple trials is crucial for retaining a high signal to noise ratio (SNR).

Since magnetic-field gradients can be relatively large inside the open shield, the sensors must be protected from vibrations or any movement, particularly those that may accompany the invoked muscle activity. Therefore, the subject rests their arm on a custom plastic mold (shown in Fig. 2a-b) which is suspended from an aluminum support that extends into the shielded region but is otherwise disconnected from the shield and sensors. The subject is thus able to make small movements of their hand within the shield without physical disturbance to the sensors and causing false signals. This was verified using control measurements in which the suspended mold and mount were moved at the expected trigger frequency without a subject arm inside.

Environmental magnetic changes in a hospital setting were measured using a fluxgate magnetometer placed outside the shield (Fig. 1e), and while large features (>100 nT on fluxgate) were visible on the OPMs, these artifacts were generally sufficiently shielded as to not cause the sensor output to go out of range during the measurement. The effects of these low-frequency transient offsets can be minimized by subtracting sensor signals (software gradiometry) and averaging.

The complete system, consisting of magnetometers, magnetic shielding, hand mounting supports, and all associated data acquisition equipment was transported by vehicle to the hospital and deployed within two hours.

Relevant photographs of the experimental equipment and setup are shown in Figure 2.

**E. Electroencephalography and electromyography**

To validate the utility of OPMs in TMS measurements, we maintain existing experimental protocols which combine TMS with EMG and EEG[12,38].

The EEG signals were recorded with a high-density (256 electrodes) EEG system (Net Station 5.0, EGI, USA). The caps were placed manually with the Cz electrode positioned over a centralized location on the scalp, which was determined as the simultaneous midpoint of the arc length for both nasion-inion and preauricular arcs. The electrode impedances were kept under 50 kOhm throughout the experiment[38-40], and a sampling frequency of 1000 Hz was used. The surface EMG was recorded from the FDI muscles. Both the EEG and EMG were digitized with a single amplifier. The amplifier applies a bandpass filter (low frequency cutoff 0.1 Hz, high frequency cutoff 70 Hz), and a notch filter (50 Hz) to the EEG. Similarly, a bandpass filter

(low frequency cutoff 0.5 Hz, high frequency cutoff 500 Hz), and a notch filter (50 Hz) were applied to the EMG.

**F. Data analysis**

The EMG-electrode data were extracted and partially analyzed using open-source Python software (MNE)[41,42]. The TMS-evoked potentials (TEP) were computed from the analysis of EEG data using Matlab 2015b and the Fieldtrip toolbox (http://www.fieldtriptoolbox.org/). The exact details of the pre-processing steps and analysis have been described elsewhere[41].

All the magnetometer data from each participant were analyzed using a custom Python code for cutting and averaging based on the TMS trigger signal. After each trigger, one second of acquired data is defined as a single trial. Since the noise within a single trial is too large (due to the magnetically hostile hospital environment) to see clear biomagnetic signals, multiple trials must be averaged together in order to achieve good signal to noise. Notch filters were applied at 50 Hz (Q=20) and higher harmonics, and the data were smoothed with an evenly weighted four-point moving window. Smoothing is preferred over low-pass filtering to avoid filter artifacts that would arise from the sharp TMS pulse.

The latency of the MEP (recorded from EMG channels) or the MEF from magnetic sensors provides important information about the nerve transmission speed. To extract latency values, a mathematical fitting function consisting of a double Gaussian with linear offset was chosen to fit the averaged data, since this would establish a repeatable and consistent way to compare the signal quality vs. the number of averages. The double Gaussian function that the data were fit to is defined as,

$$y = A_1 e^{\frac{-(x-x_1)^2}{2\sigma_1^2}} + A_2 e^{\frac{-(x-x_2)^2}{2\sigma_2^2}} + y_0 + m \times x \,,$$

where $A_1$ and $A_2$ are the individual Gaussian amplitudes, $x_1$ and $x_2$ and respective offsets, $\sigma_1$ and $\sigma_2$ are the Gaussian widths, $y_0$ is an offset and $m$ is the linear slope. The fit parameters $y_0$ and $m$ capture the decaying artifact from the TMS pulse that overlaps the MEF signal. This fit was chosen as best able to capture the bi-phasic signal and extract a consistent value for the latency between the stimulus to the onset of the action potential but has no particular physical meaning. The fit was made around 25 ms after the trigger. This offset time window was chosen to coincide with the MEF latency and to avoid fitting the artifact. The latency is then defined as $x_1 - 2.5\sigma_1$, where $x_1$ is the center offset value of the first fitted Gaussian, and $\sigma_1$ is the half-width-half-max of the fit. The value of $2.5\sigma_1$ represents a reliable point at which the data rises approximately to 5% of the Gaussian amplitude of the signal, defining a consistent value of the latency unbiased by manually chosen values. This fit was used for both the electric and magnetic data.

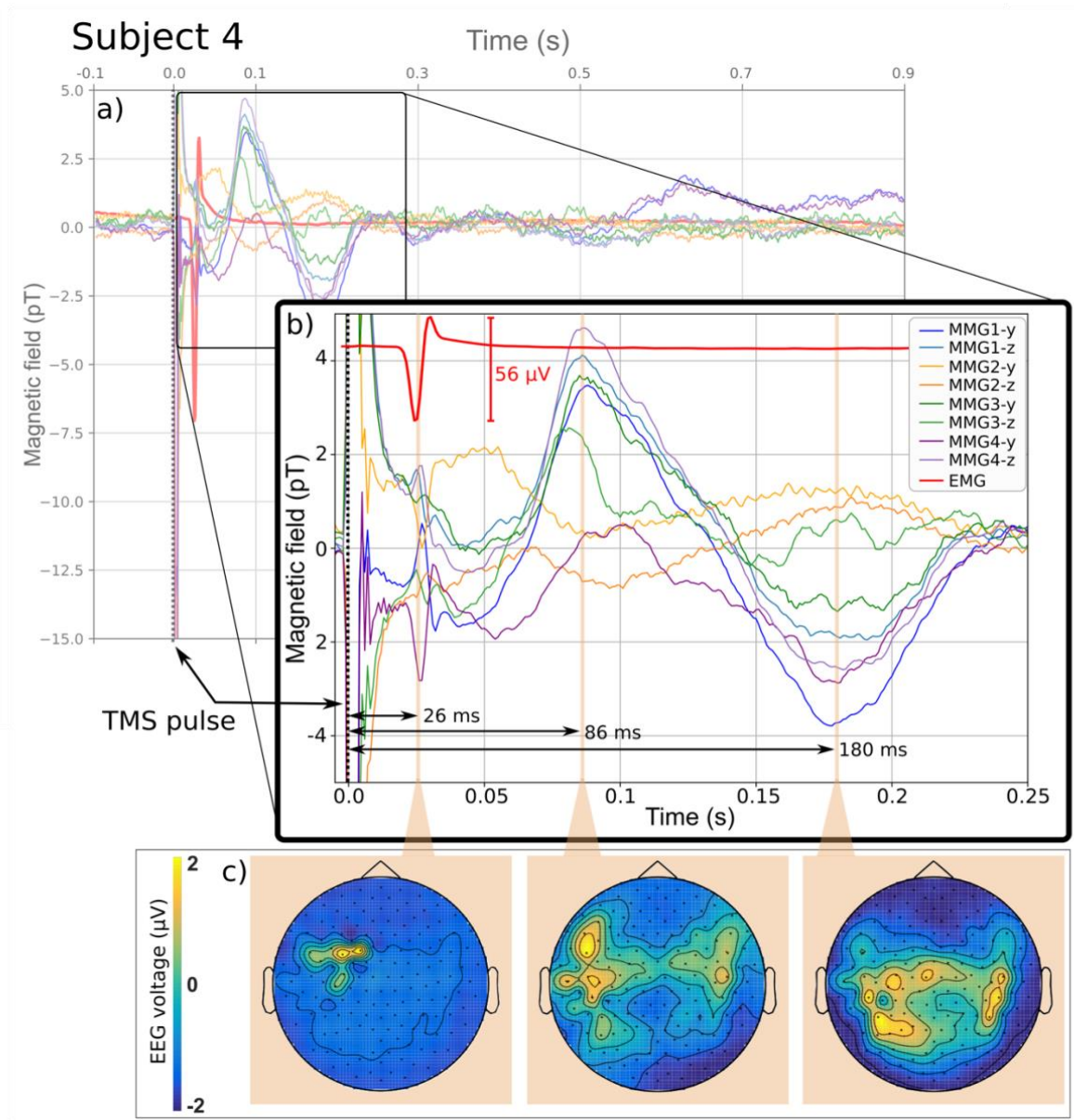

**FIG. 3**: Combined EEG, EMG and MMG data for a single participant, showing relative detail in magnetic vs. electromyography signals, and how the data from three input methods in the experimental system complement each other. (a) 120 averages of MMG and EMG data before, during and after the TMS pulse (occurring at 0.0 s). Following the large artifact at the TMS pulse, magnetic activity in the hand is detected for approximately 300 ms in this subject, which, based on control measurements, was not attributable to vibration. (b) Zoom of the data in (a) for the time period immediately following the TMS pulse. EMG channel has been offset so as not to obstruct the MMG channels. The magnetic sensors detect both activity which coincides with the electric channel, and which occurs while the electric channel shows nothing. Eight of 16 magnetic sensor channels were selected based on noise levels and signal amplitude. Referencing Fig. 1a: MMG1, magnetometer AC; MMG2, magnetometer 9P; MMG3, magnetometer AB; MMG4, magnetometer 9O. (c) EEG topograms of brain activity during selected points after the TMS pulse. The topograms were analyzed at times where magnetic features had largest amplitude. Brain activity begins in the motor cortex where the TMS pulse is applied. The activity then moves to other regions of the brain over the course of the measurement.

## III. RESULTS AND DISCUSSION

Averaged data resulting from 120 repetitive TMS pulses at 0.5 Hz with a single participant are shown in Figure 3. While 16 magnetic sensor channels are available from the experiment, we select the eight shown for clarity and consistency across subjects because some sensors failed (out of range due to environment) during measurements. The sensors shown [$y$- and $z$- axes from magnetometers MMG1(AC), MMG2(9P), MMG3(AB) and MMG4(9O)] are positioned below and above the hand, respectively. In Figure 3(a), the signals from both the EMG and MMG are shown to occur within 300 ms of the TMS pulse, with little discernible activity thereafter. During the TMS magnetic artifact, which is a homogenous field modulation over the magnetometer array, the $y$ and $z$ sensors record large features with the same sign, indicating that fields at these sensors are oriented similarly.

Figure 3(b) shows a narrower time window, where both magnetic and electric (shown in red) channels exhibit peaks at 26 ms. On the EMG channel, this feature is identified as the MEP[12], and there is good agreement in the signal latency calculated from electric and magnetic channels. The magnetic field feature associated with the MEP, which we identify as the MEF, was observed in recordings from all four subjects. On the MMG channel, the relative sign and shapes of the magnetic features in the data could be used to inform source location of the muscle activity[26].

Starting at around 50 ms after the TMS pulse, the magnetometer channels record a bi-phasic feature that lasts up to 200 ms. This larger magnetic signal does not appear on the EMG but is observable across subjects' MMG recordings. Based on control measurements, this signal is not attributable to vibrations in the system. These features may result from muscle activity arising in other parts of the hand that are not covered by the EMG electrodes, indicating that MMG can be used to map out which muscles are activated in repetitive TMS. Alternatively, the signal is consistent with an H-reflex and the loss of this H-reflex signal in EMG recording could be due to the choice of filter parameters implemented in this study[43]. Future studies will focus on identifying the source of these additional features in the magnetic signal.

The MMG data are used to identify the time points after the TMS pulse at which to examine the EEG topology plots, shown in Figure 3c. Further study is needed to establish the relationship between the MMG and EEG, but the points here are chosen as times of interest, demonstrating that data from the different signal channels are comparable. At the topology plot data at 26 ms, corresponding to the time of the MEF, electrical activity in the brain was observed at the location of the TMS stimulus, in the M1 region. At later times, 86 ms and 180 ms, this activity moved to contralateral M1 and parietal regions, respectively. Furthermore, the results showed similar patterns in other subjects (Figure 4). These results confirm that TMS induces focal effects, and these effects spread to other brain regions beyond the stimulated region[44], at timescales similar to the periphery magnetic response.

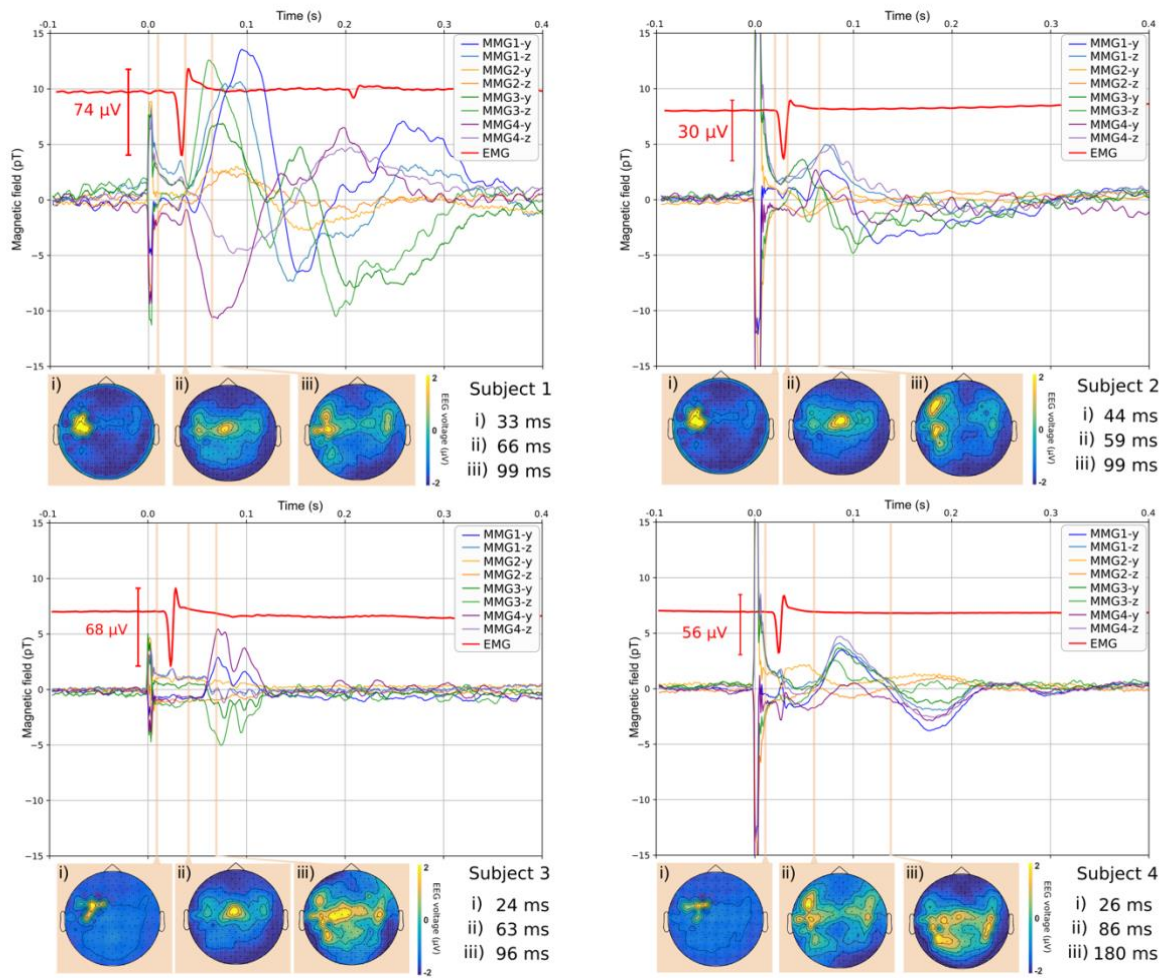

**FIG. 4:** Combined MMG, EMG, and EEG data for four participants. Variability across subjects is clearly discernible in the MMG data, but all show qualitatively similar results for all signal types. The shape of the TMS artifact is strongly dependent on the position of the TMS coil relative to the magnetic shield, which varies from participant to participant. Additionally, the actual duration of the TMS pulse is ~300 µs. The TMS artifact is distorted due to sensor bandwidth, low sampling, low-pass filtering, and timing jitter between the pulse and the sampling trigger. Note: The EMG electrode from Subject 1 was found afterwards to have been improperly grounded, leading to large noise artifacts that remained after filtering and smoothing. Nevertheless, the MEP is still visible.

Figure 4 shows the TMS-invoked magnetic and electric response of the hand for four subjects. Each subject has a unique magnetic signal – for example, data from Subject 1 shows magnetic field values that are almost three times as large as those of the other subjects. Inter-subject variability during TMS could account for variability in the EMG and MMG recordings[45]. For both EMG and MMG, variability was calculated by removing any constant or linear offsets from the signal and taking the root-mean-square value of the signal between the TMS trigger and 500 ms post trigger. The variation from the mean for each subject was calculated and the average variability for all four subjects is 28% in EMG recordings, and 34% for MMG recordings. Variability in the magnetic recordings is greater than that of the EMG. This could result from the fact that the magnetic signal is strongly dependent on the distance between source and sensor, which varies based on the subject physiology. Inter-trial variation could not be calculated because of inadequate signal-to-noise-ratio. The poor SNR arises due to operating the OPMs in the open shield, which results in a residual noise amplitude of ~10 pT before averaging.

Data from all the participants show that the MEP from the EMG is observed as a corresponding MEF in the magnetometer channels. Finally, the EEG result shows qualitatively similar behavior of the brain activity across subjects at the time points chosen based on features in the MMG.

Latency analysis demonstrates the complementarity of MMG to EMG signals in TMS measurements. There is good agreement between the latencies extracted from MMG or EMG measurements (Table 1), and the magnetic field measurement offers important validations of the electrical potential measurement. For example, EMG data can be influenced by a variety of factors involving the electrode-skin contact, including transient changes such as changing electrode impedance due to increases or decreases in skin moisture during the measurement and changing noise floors[46]. While the specific geometry of the source determines the magnetic field at the sensor, the near unity permittivity of tissue or bone means that it conveys the absolute value of the field from the source, and that it could be used to decouple changing systematic experimental conditions between measurements. In our data set, the MMG data suffers from inadequate SNR to perform such analysis on a trial-by-trial basis. Repeatable latency values could be extracted for signals averaged from at least 45 trial windows, indicating that this is the minimum number of trials needed.

**Table 1:** Comparison of MEP vs. MEF latency for each subject. Uncertainty, shown in parentheses, was calculated from covariance matrix of the fitted Gaussian function after full averaging of all trials. Uncertainty for the average is standard deviation of the four subjects. For Subject 4, there is a 1 standard-deviation discrepancy in the timing of the signal as measured from electric and magnetic channels. The good agreement in the averages indicates that there is not a statistically significant systematic over- or under- reporting of one method relative to the other.

|           | MMG [ms] | EMG [ms] |
|-----------|----------|----------|
| Subject 1 | 26(1)    | 25(1)    |
| Subject 2 | 20(2)    | 20(1)    |
| Subject 3 | 18(2)    | 19(2)    |
| Subject 4 | 22(1)    | 20(1)    |
| Average   | 22(3)    | 21(2)    |

## IV. CONCLUSION
### A. Future Work
Future work will mainly focus on increasing the SNR through implementing and optimizing active magnetic field compensation and software gradiometry. As discussed, MMG could also aid in identifying and locating TMS activated muscles that are not in regions probed by surface electrodes. In this work, the relative position between sensors and hand was not adequately controlled to perform reliable inversion of the field to acquire the source location. However, in future work, better sensor array positioning and hand position indicator methods similar to systems widely used in EEG, will be applied to achieve the high resolution widely demonstrated in the literature[47-49]. Another direction will be in combining the analysis of different system aspects, in order to probe possible connections between signals from different input channels.

### B. Significance

These first results of OPM-recorded magnetic signals from TMS-evoked movement demonstrate the future viability of the TMS-OPM system for clinical research. We showed that magnetic field sensing of periphery limbs is possible in a regular hospital using small magnetic shields, circumventing the requirement for a large and expensive magnetically shielded room, which has been a pre-requisite for previous studies. The combined use of magnetic and electric field sensors allows for detailed validation of different signals, while providing complementary information about muscle activity in the hand. TMS is targeted, repeatable and safe, and thus can be used in future studies to identify the innervation pathways for specific muscles in various locations along the arm, by using the magnetic data for magnetic source imaging.

Together with small-sized magnetic shielding, the portable and economical commercial OPM systems can enhance the utility of TMS. We demonstrated a complete MMG system that could be transported and deployed within several hours, with subject preparation times for the MMG within minutes. This approach represents a new modality in TMS research with opportunities for peripheral nerve study.

## V. ACKNOWLEDGEMENTS
The work was funded in part by the German Federal Ministry of Education and Research (BMBF) within the Quantumtechnologien program (FKZ 13N14439) and the Deutsche Forschungsgemeinschaft (DFG) through the DIP program (FO 703/2-1) and the Other Instrumentation-Based Research Infrastructure program (FKZ 324668647). S.G acknowledges support from the Transregional Collaborative Research Center (CRC) TR-128. Funding was provided by the DFG core facility, grant number KO5321/3 and TR408/11 for using the infrastructure of the core facility for ultra-low magnetic fields.

## COMPETING INTERESTS
The authors declare no competing interests.

## AUTHOR CONTRIBUTIONS
G.Z.I., Y.H. conceived of, designed, and constructed the apparatus. Y.H., G.Z.I., T.S., M.M., and V.C.C. prepared and performed the experiments. G.Z.I., Y.H., M.M., and V.C.C. analyzed the data. G.Z.I., M.M., and V.C.C. wrote the manuscript. M.M and S.G. advised and informed clinical aspects of the work. M.M, S.G., D.B., and A.W. supervised the work. All authors proofread and edited the manuscript.